\documentclass[prl,aps,letterpaper,floatfix,superscriptaddress,preprintnumbers,twocolumn,10pt]{revtex4-1}

\usepackage{dcolumn}
\usepackage{bm}
\usepackage[dvips]{graphicx}
\usepackage{epsfig}
\usepackage{amsmath, amsfonts, amssymb}

\newcommand{\be}{\begin{equation}}
\newcommand{\ee}{\end{equation}}
\newcommand{\bee}{\begin{equation*}}
\newcommand{\eee}{\end{equation*}}
\newcommand{\bea}{\begin{eqnarray}}
\newcommand{\eea}{\end{eqnarray}}
\newcommand{\bean}{\begin{eqnarray*}}
\newcommand{\eean}{\end{eqnarray*}}

\usepackage{color}

\begin{document}

\title{Looking through the Pseudo-Scalar Portal into Dark Matter: \\
Novel Mono-Higgs and Mono-$Z$ Signatures at LHC}

\author{Jose Miguel No} \email{j.m.no@sussex.ac.uk}
\affiliation{Department of Physics and Astronomy, University of Sussex, 
Brighton BN1 9QH, UK}

\date{\today}

\begin{abstract}
Mono-$X$ signatures are a powerful collider probe of the nature of dark matter. We show that mono-Higgs and mono-$Z$ may be key signatures of pseudo-scalar portal interactions 
between dark matter and the SM. We demonstrate this using a simple renormalizable version of the portal, with a Two-Higgs-Doublet-Model as electroweak symmetry breaking sector. 
Mono-$Z$ and mono-Higgs signatures in this scenario are of resonant type, which constitutes a novel type of dark matter signature at LHC. 
\end{abstract}

\maketitle

The nature of dark matter (DM) is an outstanding mystery at the interface of particle physics and cosmology. The current DM candidate paradigm is the so-called 
Weakly-Interacting-Massive-Particle (WIMP), a particle whose relic abundance is obtained via thermal freeze-out in the early Universe, and with a mass in the range 
$\mathrm{GeV}-\mathrm{TeV}$, around the scale of electroweak (EW) symmetry breaking $v = 246$ GeV. WIMP DM is very well-motivated in connection with new physics close to the 
EW scale (see \cite{Bertone:2004pz} for a review) and/or the existence of a hidden sector (singlet under the SM gauge group) which interacts with the SM 
via a portal \cite{Schabinger:2005ei,Patt:2006fw}.

A large experimental effort aims to reveal the nature of (WIMP) DM and its interactions with SM particles, either indirectly by measuring the energetic SM particles product 
of DM annihilations in space, or directly by measuring the scattering of ambient DM from heavy nuclei. Current best experimental limits on the spin-independent DM interaction 
cross section with nuclei by the Large-Underground-Xenon (LUX) experiment \cite{Akerib:2013tjd} are very strong, and particularly constraining for DM masses in the range 
$10-100$ GeV. On the other hand, the experimental limits on spin-dependent DM-nucleon interactions are much less stringent, 
generically favouring a pseudo-scalar mediator of DM-nucleon interactions (which primarily yields spin-dependent interactions) over a scalar mediator. 

Direct/indirect probes of DM are complemented by searches at colliders, where pairs of DM particles could be produced. These escape the detector and 
manifest themselves as events showing an imbalance in momentum conservation, via the presence of missing transverse momentum $E_{T}\hspace{-4mm}/$\hspace{2mm} recoiling against 
a visible final state $X$. Searches for events with large $E_{T}\hspace{-4mm}/$\hspace{2mm} are a major activity at the Large Hadron Collider (LHC) precisely 
due to their (potential) connection to DM \cite{Morrissey:2009tf}. 

Searches for DM in $X + E_{T}\hspace{-4mm}/$\hspace{2mm} channels, referred to as \textsl{mono}-$X$, can be classified according to the visible particle(s) $X$ against which the 
invisible particles recoil. Experimental studies at Tevatron and LHC have considered cases in which $X$ is a hadronic 
jet \cite{Aaltonen:2012jb,Khachatryan:2014rra,Aad:2015zva}, a photon $\gamma$ \cite{Chatrchyan:2012tea,Aad:2012fw}, $W$ or $Z$ bosons \cite{Aad:2013oja,Aad:2014vka} 
and, after the recent discovery of the Higgs boson \cite{Aad:2012tfa, Chatrchyan:2012ufa}, have also considered $X$ to be the 125 GeV Higgs particle $h$ \cite{Aad:2015yga}. 
Indeed, if DM is linked to the EW scale, $W$, $Z$ and Higgs boson signatures are natural places to search for it,
with mono-$W,Z,h$ having been recently considered as a paradigm for such potential 
signatures \cite{Petriello:2008pu,Bai:2012xg,Bell:2012rg,Carpenter:2012rg,Petrov:2013nia,Carpenter:2013xra,Berlin:2014cfa}.



With the EW symmetry breaking sector being a most natural portal to a hidden sector, it is crucial to identify key probes of such portal interactions with the DM sector. 
As pseudo-scalar portal interactions are significantly more difficult to probe experimentally via direct DM detection, collider probes constitute  
a rather unique window into these DM scenarios. In this work I investigate such probes for a renormalizable pseudo-scalar 
Higgs portal (also known as \textsl{Axion Portal}) \cite{Nomura:2008ru}, possible within extensions of the SM scalar sector.   
I show that novel mono-Higgs and mono-$W,Z$ signatures emerge in this context, with very distinct kinematical features from other mono-$X$ scenarios.
These signatures constitute a new probe of DM scenarios at LHC, deeply linked to the realization of a non-minimal Higgs sector in Nature. 

The letter is organized as follows: In Section I we introduce and discuss the pseudo-scalar portal scenario using a simple realization of a DM scenario 
in this context. In Section II we analyze the mono-Higgs and mono-$Z$ signatures the portal gives rise to, 
and discuss their prospects for the 14 TeV run of LHC. We finally outline the implications of these results in Section III.

\vspace{-3mm}

\subsection*{I. Dark Matter Through the Pseudo-Scalar Portal}

\vspace{-3mm}

The simplest realization of the pseudo-scalar portal occurs within a Two-Higgs-Doublet-Model (2HDM) extension of the SM \cite{Nomura:2008ru}.
For our purposes, we use in the following a simple embedding of DM into such a picture (see e.g. \cite{Ipek:2014gua}), and
consider dark matter to be a Dirac fermion $\psi$ with mass $m_{\psi}$, which couples to a real singlet pseudo-scalar mediator state $a_0$ via 
\begin{equation}
\label{Ldark}
 V_{\mathrm{dark}} = \frac{m^2_{a_0}}{2}\,a_0^2 + m_{\psi}\, \bar{\psi}\psi + y_{\psi}\,a_0 \,\bar{\psi} i\gamma^{5} \psi
\end{equation}
A renormalizable coupling of $a_0$ to the visible sector becomes possible by extending the SM Higgs sector to include two doublets 
$H_i = \left(\phi_i^{+} , (v_i + h_i + \eta_i)/\sqrt{2} \right)^T$ ($i=1,2$), $v_i$ being the \textit{vev} of the doublets with $\sqrt{v^2_1 + v^2_2} = v$ and 
$v_2/v_1 \equiv \mathrm{tan} \beta$. The 2HDM scalar potential reads  
\begin{eqnarray}	
\label{2HDM_potential}
V_{\mathrm{2HDM}} &= &\mu^2_1 \left|H_1\right|^2 + \mu^2_2\left|H_2\right|^2 - \mu^2\left[H_1^{\dagger}H_2+\mathrm{h.c.}\right] \nonumber \\
&+&\frac{\lambda_1}{2}\left|H_1\right|^4 +\frac{\lambda_2}{2}\left|H_2\right|^4 + \lambda_3 \left|H_1\right|^2\left|H_2\right|^2 \\
&+&\lambda_4 \left|H_1^{\dagger}H_2\right|^2+ \frac{\lambda_5}{2}\left[\left(H_1^{\dagger}H_2\right)^2+\mathrm{h.c.}\right]\nonumber 
\end{eqnarray}
where we assume Charge-Parity (CP) conservation and a $\mathcal{Z}_2$ symmetry, softly broken by $\mu$. Extending this 
$\mathcal{Z}_2$ symmetry to the couplings of the doublets $H_{1,2}$ to fermions allows to forbid potentially dangerous 
tree-level flavour changing neutral currents, by forcing each fermion type to couple to one doublet only \cite{Glashow:1976nt}. 
In Type-I 2HDM all fermions couple to $H_{2}$, while for Type-II 2HDM up-type quarks couple to $H_{2}$ and down-type quarks and leptons couple to $H_{1}$.
The pseudo-scalar portal between the visible and hidden sectors occurs via \cite{Nomura:2008ru,Ipek:2014gua}
\begin{equation}
\label{Vportal}
V_{\mathrm{portal}} = i\,\kappa\,a_0 \,H_1^{\dagger}H_2 + \mathrm{h.c.}
\end{equation}
The scalar spectrum of the 2HDM contains a charged scalar $H^{\pm} = \mathrm{cos}\beta \,\phi_2^{\pm} - \mathrm{sin}\beta \, \phi_1^{\pm}$
and two neutral CP-even scalars $h = \mathrm{cos}\alpha \,h_2 - \mathrm{sin}\alpha \, h_1$, $H_0 = - \mathrm{sin}\alpha \,h_2 - \mathrm{cos}\alpha \, h_1$.
We identify $h$ with the 125 GeV Higgs state, which is SM-like in the limit $\beta - \alpha = \pi/2$ 
(see e.g. \cite{Branco:2011iw} for a review of 2HDM). For $\kappa \neq 0$, the would-be neutral CP-odd scalar $A_0 = \mathrm{cos}\beta \,\eta_2 - \mathrm{sin}\beta \, \eta_1$ mixes 
with $a_0$ through (\ref{Vportal}), yielding two pseudo-scalar mass eigenstates $a,A$ 
\begin{equation}
A = c_{\theta} \,A_0 + s_{\theta} \, a_0 \quad, \quad a = c_{\theta} \,a_0 - s_{\theta} \, A_0
\end{equation}
with $c_{\theta} \equiv \mathrm{cos}\theta$ and $s_{\theta} \equiv \mathrm{sin}\theta$. We consider in the following the case in which the 
singlet-like mediator $a$ is lighter than $A$ ($m_A > m_a$), and $m_a > 2 \,m_{\psi}$ such that the decay $a \to \bar{\psi} \psi$ is possible. 
In terms of the mass eigenstates, the interactions (\ref{Ldark}) and (\ref{Vportal}) become 
\begin{eqnarray}	
\label{Vportal_mass}
V_{\mathrm{dark}} &\supset& y_{\psi}\,(c_{\theta} \,a + s_{\theta} \, A) \, \bar{\psi} i\gamma^{5} \psi \nonumber \\
V_{\mathrm{portal}} &=& \frac{\left( m_A^2 - m_a^2 \right) s_{2\theta}}{2\,v} \, \left(c_{\beta-\alpha}\,H_0 - s_{\beta-\alpha}\,h \right)  \\
&\times& \left[ a A \,(s_{\theta}^2 - c_{\theta}^2) + (a^2 - A^2)\, s_{\theta} c_{\theta}  \right] \, . \nonumber 
\end{eqnarray}
Gauge interactions of the two doublets $H_i$ yield the relevant interactions $aZh \propto s_{\theta}\,c_{\beta-\alpha}$, 
$AZh \propto c_{\theta}\,c_{\beta-\alpha}$, $aZH_0 \propto s_{\theta}\,s_{\beta-\alpha}$, $AZH_0 \propto c_{\theta}\,s_{\beta-\alpha}$, 
$aW^{\pm}H^{\mp} \propto s_{\theta}\,s_{\beta-\alpha}$, $AW^{\pm}H^{\mp} \propto c_{\theta}\,s_{\beta-\alpha}$, while 
$V = V_{\mathrm{2HDM}} + V_{\mathrm{dark}} + V_{\mathrm{portal}}$ yields 
$aAh \propto s_{4\theta}\, s_{\beta-\alpha}$, $a\bar{\psi}\psi \propto c_{\theta}$ and 
$A\bar{\psi}\psi \propto s_{\theta}$.

\vspace{2mm}

Altogether, the interactions above lead to mono-$h$ and mono-$W,Z$ signatures at LHC in various possible ways, which we discuss in detail in the next Section. 
In particular, for $m_A > m_h + m_a$, $m_{H_0} > m_Z + m_a$, $m_{H^{\pm}} > m_{W^{\pm}} + m_a$, 
this scenario yields a novel signature: ``\textsl{resonant mono-}$h,W,Z$" respectively via the processes 
$p p \to A \to h\,a$, $p p \to H_0 \to Z\,a$, $p p \to H^{\pm} \to W^{\pm}\,a$, with the mediator $a$ subsequently decaying into DM. 

\subsection*{II. Mono-Higgs and Mono-$W,Z$ Signatures at LHC}

\vspace{-3mm}

As outlined above, in these scenarios there are two different kinds of processes which, through the production of $X + \bar{\psi} \psi$ ($X = W,Z,h$), lead to 
mono-$X$ signatures at LHC. Focusing on mono-Higgs for the purpose of illustration, there exist contributions from 
$p p \, (\bar{q}q) \to Z^* \to h\,a \,(a \to \bar{\psi}\, \psi)$ and $p p\, (g g) \to A \to h\,a \,(a \to \bar{\psi} \,\psi)$.

The former is kinematically similar to mono-$h$ signatures in other scenarios \cite{Petrov:2013nia,Carpenter:2013xra,Berlin:2014cfa}, 
which are generically suppressed either by the presence of an off-shell or very massive particle in the s-channel.
Together with the momentum transfer being cut-off by the parton distribution functions (PDFs), this leads to very small mono-Higgs cross sections, 
making a mono-$h$ signature difficult to probe at the 14 TeV run of LHC even with a large integrated luminosity, 
if it solely arises from this type of contribution.

In contrast, for $m_A > m_h + m_a$ the kinematics of the latter process is very different, due to $A$ being resonantly produced. 
In this case, the 4-momentum of $h$ and $a$ is kinematically fixed, and $E_{T}\hspace{-4mm}/$\hspace{2mm} is bounded from above by 
\begin{equation}
\label{ETmax}
E^{\mathrm{max}}_{T}\hspace{-7.5mm}/ \hspace{5.5mm}= \frac{1}{2\, m_{A}} \sqrt{(m^2_A - m^2_h - m^2_a)^2 - 4 \,m^2_h\,m^2_a}\, .
\end{equation}
The $E_{T}\hspace{-4mm}/$\hspace{2mm} distribution from this process is a steeply rising function 
with a sharp cut-off at $E^{\mathrm{max}}_{T}\hspace{-7.5mm}/$\hspace{5.5mm}, a very distinct feature of these scenarios.
At the same time, this contribution to mono-$h$ is resonantly enhanced \textit{w.r.t} the former one, generically yielding a much larger cross section.
Furthermore, it is important to stress that in this scenario the resonant contribution is proportional to $s_{\beta-\alpha}^2$ and thus maximal in the 
2HDM alignment limit of a SM-like Higgs $h$ (as favoured by ATLAS and CMS analyses), whereas the off-shell contribution is proportional to 
$c_{\beta-\alpha}^2$, vanishing in that limit. 

Before we continue, let us briefly comment on the fact that such resonant mono-$h$ signatures may also occur in a pure 2HDM 
through the process $p p \,(g g) \to A \to h\,Z \,(Z \to \nu \nu)$. We stress that the phenomenology in the presence of the 
pseudo-scalar portal to DM is radically different from that of the pure 2HDM. First, contrary to the case of the DM portal,
the interaction yielding a mono-$h$ signature in the 2HDM vanishes for a SM-like Higgs $h$, as discussed above. Second and most important,
for a pure 2HDM the same process with $Z \to \ell \ell$ is a much more sensitive probe of the existence of $A$ than the mono-$h$ signature.  
This constitutes a generic, crucial way of disentangling a \textsl{resonant} $X + E_{T}\hspace{-4mm}/$\hspace{2mm} signature 
where $E_{T}\hspace{-4mm}/$\hspace{2mm} originates in a dark sector (\textsl{e.g.} $a \to \bar{\psi}\psi$) from that where 
the $E_{T}\hspace{-4mm}/$\hspace{2mm} comes from $Z \to \nu \nu$, as the latter will have to be accompanied by a much more sensitive   
$Z \to \ell \ell$ counterpart, while the former will not.

For our phenomenological analysis, we choose a Type II 2HDM benchmark $t_{\beta} = 3$, $c_{\beta-\alpha} = 0.05$ (close to the 2HDM alignment limit) with 
$s_{\theta} = 0.3$, corresponding to a moderate mixing between the visible and dark sectors, and $y_{\psi} = 0.2$. For the mediator and DM masses 
we choose respectively $m_a = 80$ GeV, $m_{\psi} = 30$ GeV. The DM annihilation cross section in this case is of order needed for a correct
DM relic density \cite{Ipek:2014gua}, and we find that for $m_a$, $m_{\psi}$ masses in this ballpark, $a \to \bar{\psi}\psi$
yields the dominant branching fraction for $y_{\psi} \gtrsim 0.02$, and $\mathrm{BR}(a \to \bar{\psi} \psi) > 0.99$ for $y_{\psi} \gtrsim 0.1$.
We take for simplicity  $m_{H^{\pm}} = m_{H_0} = m_{A}$, verifying that it satisfies EW precision observable constraints, 
and for each value of $m_{A}$ we adjust $\mu$ in (\ref{2HDM_potential}) to be within the region compatible with vacuum stability, 
perturbativity and unitarity.

\begin{figure}[t]
\begin{center}
\includegraphics[width=0.48\textwidth]{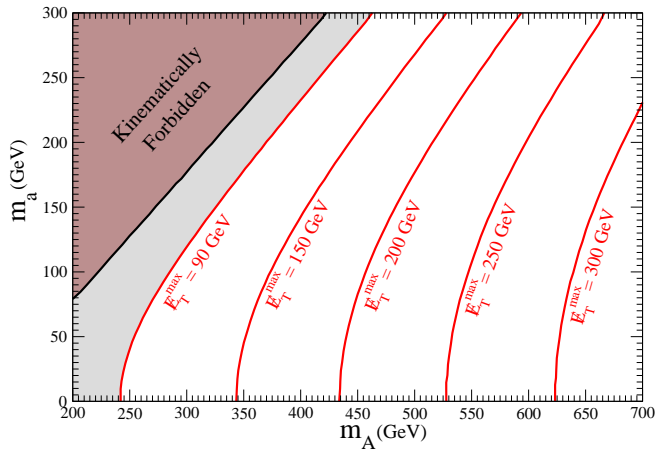}
\caption{\small Value of $E^{\mathrm{max}}_{T}\hspace{-7.0mm}/$\hspace{5.5mm} from (\ref{ETmax}) for resonant mono-$h$ in the ($m_A,\,m_a$) plane. 
In the solid-brown region the decay $A \to h\, a$ is kinematically forbidden. The grey region lies below the event selection of \cite{Aad:2015yga}. 
A similar Figure may be obtained in the place ($m_{H_0},\,m_a$) for resonant mono-$Z$.}
\label{fig:1}
\end{center}
\vspace{-3mm}
\end{figure}

In the following, taking as benchmark values $m_{H_0} = m_{A} = 300,\,500,\,700$ GeV, which we denote respectively as benchmarks A, B, C, 
we discuss the existing bounds from the 8 TeV LHC run and explore the 14 TeV LHC run prospects for \textsl{resonant mono}-$h,Z$.

\vspace{-3mm}

\subsubsection{Mono-Higgs}

\vspace{-3mm}

Current ATLAS and CMS mono-$h$ searches focus on the $h \to \gamma \gamma$ decay of the 125 GeV Higgs boson. 
For our analysis we use the selection criteria from the LHC 8 TeV run data analysis by ATLAS \cite{Aad:2015yga},
which selects events with two photons with leading (subleading) transverse momentum $P^{\gamma}_T > 35$ ($25$) GeV, rapidity $|\eta^{\gamma}| < 2.37$ 
and in the invariant mass window $m_{\gamma\gamma} \in \left[105,160\right]$ GeV. In addition, the photon pair is required 
to have been produced in association with a sizable amount of missing transverse momentum, $E_{T}\hspace{-4mm}/\hspace{2mm} > 90$ GeV, and such that 
$P^{\gamma\gamma}_{T} > 90$ GeV (to suppress background events where $E_{T}\hspace{-4mm}/$\hspace{2mm} is caused by 
mismeasurement of energies of identified physical objects). ATLAS yields a 95 \% C.L. upper bound on the 
cross section of $0.70$ fb, while our 8 TeV signal samples for $m_{A} = 300,\,500,\,700$ GeV generated with 
{\sc MadGraph5$\_$aMC$@$NLO} \cite{Alwall:2011uj,Alwall:2014hca} respectively yield, after selection cuts, 
0.143 fb, 0.043 fb and 0.011 fb, including next-to-leading-order (NLO) QCD effects computed using {\sc Sushi} \cite{Harlander:2012pb}.

In Figure \ref{fig:1} we show the value of $E^{\mathrm{max}}_{T}\hspace{-7.5mm}/$\hspace{5.5mm} for resonant mono-$h$ in the ($m_A,\,m_a$) plane, which 
highlights the fact that, while current searches are not sensitive to $m_{A} \lesssim 250$ GeV (as $E^{\mathrm{max}}_{T}\hspace{-7.5mm}/\hspace{5.5mm} < 90$ GeV),
the value of $E^{\mathrm{max}}_{T}\hspace{-7.5mm}/$\hspace{5.5mm} rapidly increases with $m_A$, making the signature 
$p p \to h\,\bar{\psi}\psi\,\, (h\to \gamma\gamma)$ promising for masses $m_{A} \gtrsim 300$ GeV at the LHC 14 TeV run.
For our analysis of resonant mono-$h$ prospects at LHC 14 TeV, we generate our signal and background event
samples with {\sc MadGraph5$\_$aMC$@$NLO}. These are passed on to {\sc Pythia} \cite{Sjostrand:2007gs} for parton showering and hadronization, 
and then to {\sc Delphes} \cite{deFavereau:2013fsa} for a detector simulation. 
The main SM backgrounds are $Z\gamma\gamma$, $Z\gamma$+jets (with a jet being misidentified as a photon, the fake rate being 
$P_{j\to\gamma} \sim 10^{-3}$ \cite{Lilley:2011mea})
and SM Higgs associated production $Z h$, with $Z \to \nu \nu$. Backgrounds with a $W$ instead of a $Z$ boson may be suppressed by vetoing extra leptons. 
NLO cross section values are estimated via $K$-factors: $K\simeq 1.65, \, 1.3$ respectively for $Z\gamma\gamma$ and $Z\gamma$+jets \cite{Campbell:2012ft},
$K\simeq 1.3$ for $Z h$ \cite{Maltoni:2013sma} and $K\simeq 2.27, \, 1.8, \,1.69$ respectively for our signal benchmarks A, B, C through {\sc Sushi}. 

\begin{table}[ht]
\begin{tabular}{l| c | c | c | c | c | c |}

& A & B & C & $Z h$  & $Z\gamma\gamma$ & $Z\gamma\,j$ \\
\hline 
&  &  &  &  & &\\ [-2ex]
Event selection& 249 & 56 & 16 & 51 & 517 & 157\\ [0.5ex]
$m_{\gamma\gamma} \in [120,130]$ GeV& 161 & 26 & 6 & 34 & 97 & 32\\ [0.5ex]
$E_{T}\hspace{-4mm}/\hspace{2mm},\,P^{\gamma\gamma}_{T} > 80$ GeV& 105 & 24 & 5 & 13 & 32 & 12\\ [0.5ex]
$E_{T}\hspace{-4mm}/\hspace{2mm},\,P^{\gamma\gamma}_{T} > 180$ GeV& 4 & 15 & 4 & 2 & 3 & 2\\ [0.5ex]
$E_{T}\hspace{-4mm}/\hspace{2mm},\,P^{\gamma\gamma}_{T} > 280$ GeV& $<$ 0.1 & 2 & 3 & 0.4 & 0.5 & 0.5\\ [0.5ex]
\hline
\end{tabular}
\caption{\small Expected number of events after event selection (see text for details) and signal region cuts for mono-$h$ with $h \to \gamma \gamma$, for 
LHC 14 TeV with $\mathcal{L} = 300\,\,\mathrm{fb}^{-1}$. Signal benchmarks A, B, C are described in Section II.}
\label{Table1}
\end{table}

In Table \ref{Table1} we show the expected signal and background events for LHC at 14 TeV with an integrated luminosity $\mathcal{L} = 300\,\,\mathrm{fb}^{-1}$,  
after event selection and in the signal region. Event selection requirements for the two photon candidates 
follow \cite{Aad:2015yga} and are described above, dropping the $E_{T}\hspace{-4mm}/\hspace{2mm} > 90$ GeV cut. We subsequently define the signal region via 
$m_{\gamma\gamma} \in [120,130]$ GeV and $E_{T}\hspace{-4mm}/\hspace{2mm},\,P^{\gamma\gamma}_{T} > 80$ GeV, $180$ GeV, $280$ GeV respectively to maximixe the sensitivity 
to benchmarks A, B, C. The $E_{T}\hspace{-4mm}/$\hspace{2mm} distribution for the three signal benchmarks and main backgrounds 
before applying this last cut is shown in Figure \ref{fig:2} (LEFT). From Table \ref{Table1}, we see that, upon neglecting systematic uncertainties, 
an approximate significance $S/\sqrt{S+B} \sim 7.9, 3.2, 1.5$ is obtained in the signal region respectively for benchmarks A, B, C and $\mathcal{L} = 300\,\,\mathrm{fb}^{-1}$.

\vspace{-3mm}

\subsubsection{Mono-Z}

\vspace{-3mm}

The recent ATLAS search \cite{Aad:2014vka} constraints mono-$Z$ signatures with $Z \to \ell^+ \ell^-$ using the available LHC 8 TeV run data. 
Their analysis selects events with two opposite sign (opposite charges) electrons/muons in the invariant mass window $m_{\ell\ell} \in \left[76,106\right]$ GeV, 
with $P^{\ell}_T > 20$ GeV and rapidity $|\eta^{\ell}| < 2.5$ ($2.47$) for muons (electrons). The rapidity of the di-lepton system has to satisfy 
$|\eta^{\ell\ell}| < 2.5$, and event selection further requires
\begin{equation}
\Delta\phi(\vec{E_{T}}\hspace{-4mm}/\hspace{2mm}, \vec{P_T^{\ell\ell}}) > 2.5\,, \quad |P_T^{\ell\ell} - E_{T}\hspace{-4mm}/\hspace{2mm}|/P_T^{\ell\ell} < 0.5\,.
\end{equation}

Four signal regions are defined, correponding respectively to $E_{T}\hspace{-4mm}/\hspace{2mm} > 150$ GeV, $250$ GeV, $350$ GeV and $450$ GeV. The ATLAS
analysis yields respective 95 \% C.L. observed upper bound on the cross section of $2.7$ fb, $0.57$ fb, $0.27$ fb and $0.26$ fb. 
Our three signal benchmark scenarios, A, B, C, satisfy these bounds, and as we show in the following they are very promising for the 14 TeV run of LHC.

\begin{widetext}
\onecolumngrid

\begin{figure}[ht!]
\begin{center}
\includegraphics[width=0.52\textwidth]{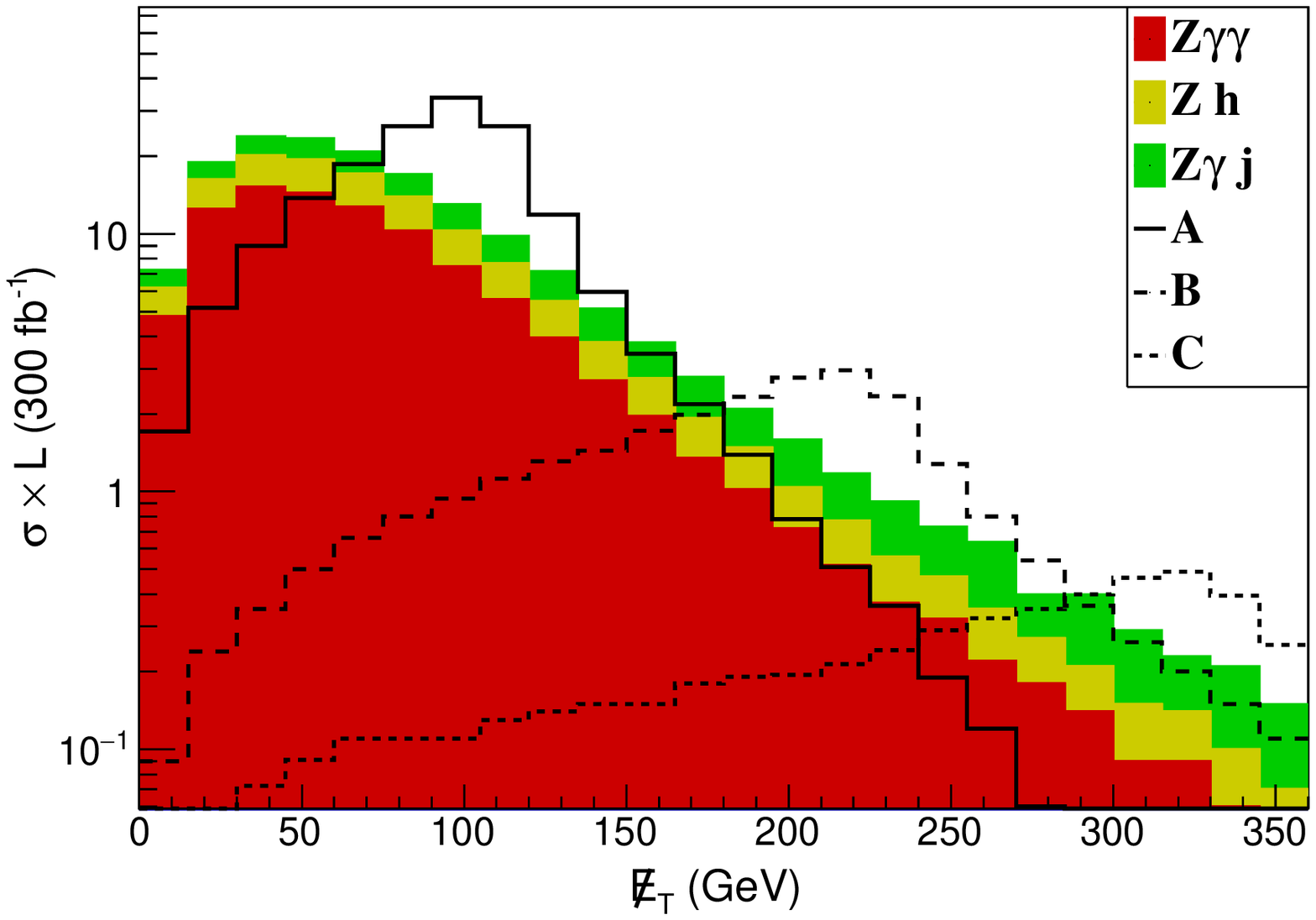}\hspace{-10mm}
\includegraphics[width=0.52\textwidth]{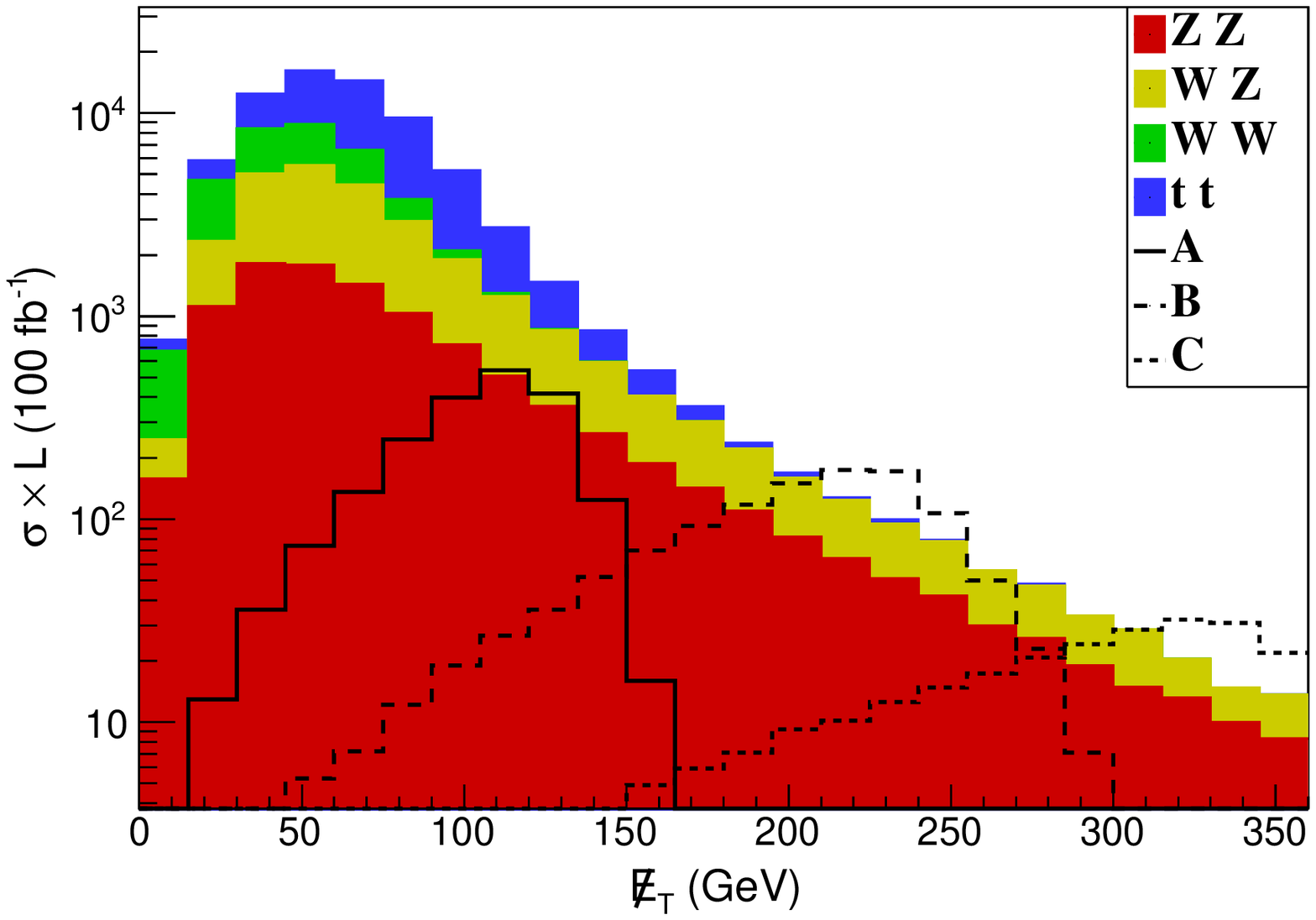}
\caption{\small LEFT: $E_{T}\hspace{-4mm}/$\hspace{2mm} distribution for mono-$h$ signal benchmarks A (solid black), B (dashed black), C (fine-dashed black) 
and background processes $Z \gamma \gamma$ (red), $Z h$ (yellow) and $Z \gamma j$ (green), yielding $E_{T}\hspace{-4mm}/$\hspace{2mm}$\,\,+\,\, \gamma\gamma$, 
after event selection (see text for details) and for $m_{\gamma\gamma} \in \left[120,130\right]$ GeV.
RIGHT: $E_{T}\hspace{-4mm}/$\hspace{2mm} distribution for mono-$Z$ signal benchmarks A, B, C 
and background processes $Z Z$ (red), $W Z$ (yellow), $W W$ (green) and $t \bar{t}$ (blue), yielding $E_{T}\hspace{-4mm}/$\hspace{2mm}$\,\,+\,\, \ell^+\ell^-$, 
after event selection (see text for details). In both cases, backgrounds are stacked on top of each other while signals are not, with bins being 
$15$ GeV wide and normalized to show the number of events per bin.}
\label{fig:2}
\end{center}
\end{figure}

\end{widetext}

For our resonant mono-$Z$ analysis at LHC 14 TeV, we follow a similar procedure to the one described for the mono-$h$ case in the previous section,
using {\sc MadGraph5$\_$aMC$@$NLO}, {\sc Pythia} and {\sc Delphes} for our signal $p p \to Z\, a$ ($Z \to \ell^+ \ell^-$, $a \to \psi \psi$) 
and background event samples. The SM irreducible backgrounds 
are $Z Z \to \ell^+ \ell^-\, \nu \bar{\nu}$ and $W W \to \ell^+ \nu\, \ell^- \bar{\nu}$, while $W Z \to \ell \nu\,\ell^+ \ell^- $ and 
$t \bar{t} \to b \ell^+ \nu \,\bar{b} \ell^- \bar{\nu}$ are the most important reducible backgrounds.
NLO cross sections are estimated via $K$-factors: $K\simeq 1.2, 1.79, 1.68$ respectively for $Z Z, W Z$ and $W W$ \cite{Ohnemus:1994ff},
$K\simeq 1.5$ for $t \bar{t}$ \cite{Maltoni:2013sma} and $K\simeq 2.36, \, 1.88, \,1.75$ respectively for our signal benchmarks A, B, C via {\sc Sushi}. 
Our event selection follows \cite{Aad:2014vka} and is discussed above, and we define three signal regions 
$E_{T}\hspace{-4mm}/\hspace{2mm},\,P^{\gamma\gamma}_{T} > 90$ GeV, $190$ GeV, $290$ GeV to respectively maximixe sensitivity 
to benchmarks A, B, C.

In Table \ref{Table2} we show the expected signal and background events for LHC at 14 TeV with an integrated luminosity $\mathcal{L} = 100\,\,\mathrm{fb}^{-1}$,  
after event selection and in the various signal regions. Neglecting systematic uncertainties, 
an approximate significance $S/\sqrt{S+B} \sim 12.8, 18.7, 9.2$ is obtained in the respective optimal signal region for benchmarks A, B, C.
In Figure \ref{fig:2} (RIGHT), we show the $E_{T}\hspace{-4mm}/$\hspace{2mm} distribution for signal and background after event selection. 

\begin{table}[ht]
\begin{tabular}{l| c | c | c | c | c | c | c|}

& A & B & C & $Z Z$  & $W W$  &$W Z$ & $t \bar{t}$ \\
\hline 
&  &  &  &  & &  &\\ [-2ex]
Event selection& 2009 & 1130 & 282 & 10100 & 12670 & 16680 & 32060 \\ [0.5ex]
$E_{T}\hspace{-4mm}/\hspace{2mm} > 90$ GeV& 1500 & 1105 & 279 & 2660 & 253 & 3530 & 5660\\ [0.5ex]
$E_{T}\hspace{-4mm}/\hspace{2mm} > 190$ GeV& 4.5 & 733 & 254 & 414 & $<$ 0.1 & 357 & 30\\ [0.5ex]
$E_{T}\hspace{-4mm}/\hspace{2mm} > 290$ GeV& 1.5 & 11 & 158  & 81 & - & 57 & $<$ 0.1\\ [0.5ex]
\hline
\end{tabular}
\caption{\small Expected number of events after event selection (see text for details) and in the signal region for mono-$Z$ with $Z \to \ell^+ \ell^-$, for 
LHC 14 TeV with $\mathcal{L} = 100\,\,\mathrm{fb}^{-1}$. Signal benchmarks A, B, C are described in Section II.}
\label{Table2}
\end{table}

Finally, although not discussed in his work, resonant mono-$W$ signatures are also possible in this setup, 
but the suppressed production of $H^{\pm}$ compared to $A$/$H_0$ makes them much less promising. 

\subsection{III. Discussion and Outlook}

\vspace{-3mm}

The analysis of the previous Section shows that resonant mono-Higgs and mono-$Z$ are promising signatures for the 14 TeV run of LHC 
with $\mathcal{L} = 100 - 300 \,\,\mathrm{fb}^{-1}$, with mono-$Z$ in particular being a very sensitive probe of pseudo-scalar portal scenarios like the one discussed here. 
Moreover, not only these signatures constitute a window into the DM sector, but are also potential discovery modes for the heavy states of the non-minimal scalar sector 
(here $A, H_0$), as their ``usual" decay modes (e.g. in a pure 2HDM) will get suppressed by the presence of the new decay channels
into the dark sector. 

Finally, there are other possible avenues for exploring these pseudo-scalar portal scenarios, like mono-jet searches on $p p \,(g g) > a \,j\, (a \to \bar{\psi} \psi)$, 
which we do not explore here, but will most certainly be complementary to the ones introduced in this work.
One other aspect I have not explored is the possibility of a light $a$ state, such that $m_h - m_Z  > m_a > 2\, m_{\psi}$. The 
exotic Higgs decay $h \to Z \,a \to \ell^+ \ell^- \, \bar{\psi} \psi$ is an interesting signature of such scenarios, and will be  
studied elsewhere.

\vspace{2mm}

To conclude, I have shown that, for DM scenarios with a pseudo-scalar mediator between the visible and DM sectors, novel potential
collider signatures emerge in the form of \textsl{resonant mono-}$h,Z$. These constitute a new probe of DM scenarios at LHC, very promising for the 
14 TeV run, and would also point to the realization of a non-minimal Higgs sector in Nature.

\begin{center}
\textbf{Acknowledgements} 
\end{center}

\vspace{-2mm}

I want to thank the organizers of the Les Houches 2015 Workshop ``Physics at TeV Colliders" where this work begun, 
and also Ken Mimasu, Veronica Sanz, Seyda Ipek and Belen Lopez-Laguna for very useful discussions.
J.M.N. is supported by the People Programme (Marie curie Actions) of the European Union Seventh Framework Programme (FP7/2007-2013) under REA grant agreement PIEF-GA-2013-625809.


\end{document}